\documentclass[aps,prd,amsmath,twocolumn,groupedaddress,eqsecnum,showpacs,showkeys,nofootinbib,preprintnumbers]{revtex4}

 \newcommand{\LT}{Lema\^{\i}tre-Tolman}

\newcommand{\df}{\ {\overset {\rm def} =}\ }

 \renewcommand{\d}{{\rm d}}     
 \newcommand{\p}{{\partial}}
 \newcommand{\td}[2]{\frac{{\rm d} {#1}}{{\rm d} {#2}}}

 \newcommand{\pdil}[2]{\partial {#1} / \partial {#2}}
 \newcommand{\nn}{\nonumber}

 \usepackage[dvips]{graphicx}
 \usepackage{amsmath}
 \usepackage{amsfonts}
 \usepackage{amssymb}

\begin{document}

 \title{
 Alternative Methods of Describing Structure Formation in the {\LT} Model}

 \author{Charles Hellaby}
 \affiliation{Department of Mathematics and Applied Mathematics, \\
 University of Cape Town, Rondebosch 7701, South Africa}
 \email{cwh@maths.uct.ac.za}

 \author{Andrzej Krasi\'{n}ski}
 \affiliation{N. Copernicus Astronomical Center,
 Polish Academy of Sciences, \\
 Bartycka 18, 00 716 Warszawa, Poland}
 \email{akr@camk.edu.pl}

 \date {11/8/2005}



 \begin{abstract}
   We describe several new ways of specifying the behaviour of 
 {\LT} (LT) models, in each case presenting the method for 
obtaining the LT arbitrary functions from the given data, and the conditions 
for existence of such solutions.  In addition to our previously considered 
`boundary conditions', the new ones include: a simultaneous big bang, a homogeneous 
density or velocity distribution in the asymptotic future, a simultaneous big 
crunch, a simultaneous time of maximal expansion, a chosen density or velocity 
distribution in the asymptotic future, only growing or only decaying fluctuations.  
Since these conditions are combined in pairs to specify a particular model, this 
considerably increases the possible ways of designing LT models with desired 
properties.
 \end{abstract}

 \maketitle

 \section{Aim and Motivation}

The {\LT} (LT) metric is the most widely used model of
cosmic inhomogeneity, being suitable for both large and small scale
inhomogeneities.  As the simplest inhomogeneous solution of Einstein's
equations, it is relatively easy to work with, but retains the full
nonlinearity of the field equations.

   However, because of this nonlinearity, generating models with a
specific desired evolution was not all that easy, and involved a large
measure of guesswork in chosing the arbitrary functions of the model, as
well as repeated numerical evolution to check if satisfactory results had
been obtained.

The usefulness of the LT metric is greatly increased if specific models can be
designed to have certain properties that satisfy observational or theoretical
requirements.  Previous papers \cite{KrHe2002,KrHe2003a,KrHe2003b}, (hereafter
papers I, II and III) have shown how to generate LT models that evolve from a
given (spherically symmetric) density or velocity profile at time $t_1$ to a
second given density or velocity profile at time $t_2$.

   We here extend the methods by which models may be constructed.  We seek
to solve for the arbitrary functions $E(M)$ and $t_B(M)$ that determine
the LT model that evolves according to the following requirements, in
various combinations, where the first two were considered previously:
 \begin{itemize}
 \item   a density profile $\rho_i(M)$ is given at time $t_i$,
 \item   a velocity profile $(R_{,t})_i(M)$ is given at time $t_i$,
 \item   the bang time is simultaneous,
 \item   the cruch time is simultaneous,
 \item   the time of maximum expansion is simultaneous,
 \item   the model becomes homogeneous at late times,
 \item   only growing modes are present,
 \item   only decaying modes are present,
 \item   a velocity profile $(R_{,t})(M)$ is given at late times,
 \item   a
 time-scaled density profile $t^3 \rho(M)$ is given at late times.
 \end{itemize}
 It requires two of the above conditions to specify an LT model, and in 
various contexts different combinations may be useful.  We work them out 
here in a systematic way for future reference.  Where two conditions are 
specified on the same time slice, this often requires special treatment.  
(In contrast, \cite{MuHeEl97} shows how to determine the LT functions from 
observational data on the past null cone.)

 \section{Outline of the {\LT} model}

   The {\LT} (LT) model \cite{Lema1933, Tolm1934} is a
spherically symmetric, nonstatic solution of the Einstein equations that
is inhomogeneous in the radial direction.  The matter source is a perfect
fluid with zero pressure, i.e. dust, and the coordinates are comoving with
the matter particles.  The metric is
 \begin{align}
   \d s^2 = \d t^2 - \frac {{R_{,r}}^2}{1 + 2E(r)} \d r^2 -
   R^2(t,r)(\d\vartheta^2 + \sin^2\vartheta\d\varphi^2) ,
 \end{align}
where $E(r)$ is an arbitrary function that determines the local
curvature of constant $t$ slices, and $R_{,r}$ is the $r$ derivative of
the areal radius $R(t,r)$.

   With $\Lambda = 0$ assumed, the evolution equation for $R$ is
 \begin{align}
   {R_{,t}}^2 = \frac{2M}{R} + 2 E .
   \label{Rtsq}
 \end{align}
 where $M(r)$ is a second arbitrary function, representing the total
gravitational mass within the comoving matter shell at constant $r$.
Note that in this equation $E(r)$ has a second interpretation as the
local energy per unit mass of the dust particles, and thus it also
determines the type of evolution (see below).  The density is
 \begin{align}
   \kappa \rho = \frac {2M_{,r}}{R^2R_{,r}}
   \qquad \text{where} \quad \kappa = \frac {8\pi G} {c^4} .
   \label{rhoLT}
 \end{align}
 The evolution equation (\ref{Rtsq}) has three types of solution: \\
 Elliptic, $E < 0$:
 \begin{multline}
   R(t,r) = \frac{M}{(-2E)}(1 - \cos\eta) , \\
   \eta - \sin\eta  = \frac {(-2E)^{3/2}}{M} (t - t_B(r)) ;
   \label{EllEv}
 \end{multline}
 Parabolic, $E = 0$:
 \begin{align}
   R(t,r) = \left[ \frac{9}{2} M (t - t_B(r))^2\right]^{1/3} ;
   \label{ParEv}
 \end{align}
 Hyperbolic, $E > 0$:
 \begin{multline}
   R(t,r) = \frac{M}{2E}(\cosh\eta - 1) , \\
   \sinh\eta - \eta = \frac {(2E)^{(3/2)}}{M} (t - t_B(r)) ;
   \label{HypEv}
 \end{multline}
 where $\eta$ is a parameter and $t_B(r)$ is the third arbitrary function, 
representing the local time at which the big bang occurs.  
The parabolic evolution is the $E \to 0$
limit of the other two evolutions, obtained by noting that $\eta/\sqrt{-2E}$
remains finite.  It is perfectly possible to have adjacent elliptic and
hyperbolic regions in one model, the evolution being parabolic on the boundary
where $E = 0$, but in general $E' \neq  0$.

   Elliptic models have both a big bang $t = t_B$ and a big crunch,
 \begin{align}
   t_C(r) = t_B + T(r) ,
 \end{align}
 where the lifetime of each worldline is
 \begin{align}
   T(r) = \frac{2 \pi M}{(-2E)^{3/2}}.
   \label{Lifetime}
 \end{align}
 The instant of maximum expansion is
 \begin{align}
   t_{MX}(r) = t_B + \frac{\pi M}{(-2E)^{3/2}},
   \label{t.MX}
 \end{align}
 at which moment the maximum areal radius
 \begin{align}
   R_{MX}(r) = \frac{M}{(-E)}
   \label{R.MX}
 \end{align}
 is reached.

   The homogeneous case is obtained by setting
 \begin{align}
   E \propto M^{2/3} , \qquad t_B = \text{constant}
 \end{align}
and it is a Friedmann model, i.e. a Robertson-Walker (RW) model with
zero pressure.

 In the following, we will use the notation
 \begin{align}
   a & = R/M^{1/3} ,
      \label{aiDef} \\
   x & = (-2E)/M^{2/3} , && \qquad E < 0 ,
      \label{xDefE} \\
   x & = (2E)/M^{2/3} , && \qquad E \geq 0 ,
      \label{xDefHP} \\
   b & = R_{,t}/M^{1/3} .
   \label{biDef}
 \end{align}
 The parametric solutions (\ref{EllEv}) and (\ref{HypEv}) may be written,
for the expanding hyperbolic (HX), expanding elliptic (EX), collapsing
elliptic (EC), and collapsing hyperbolic (HC) cases as
 \begin{widetext}
 \begin{align}
   \text{HX:} \quad t & = t_B + x^{-3/2} \left\{
      \sqrt{(1 + x a)^2 - 1}\; - {\rm arcosh} (1 + x a)
      \right\} , \quad && E > 0 , \quad t > t_B ;
      \label{t.of.R.Hyp.exp} \\
   \text{EX:} \quad t & = t_B + x^{-3/2} \left\{
      \arccos ( 1 + x a) - \sqrt{1 - (1 + x a)^2}\;
      \right\} , \quad && E < 0 , \quad 0 \leq \eta \leq \pi ;
      \label{t.of.R.Ell.exp} \\
   \text{EC:} \quad t & = t_B + x^{-3/2}  \left\{
      2 \pi - \arccos (1 + x a) + \sqrt{1 - (1 + x a)^2}\;
      \right\} , \quad && E < 0 , \quad \pi \leq \eta \leq 2 \pi ;
      \label{t.of.R.Ell.col} \\
   \text{HC:} \quad t & = t_C - x^{-3/2}  \left\{
      \sqrt{(1 + x a)^2 - 1}\; - {\rm arcosh} (1 + x a)
      \right\} , \quad && E > 0 , \quad t < t_C .
      \label{t.of.R.Hyp.col}
 \end{align}
 Alternatively, from (\ref{Rtsq}) and (\ref{xDefE}) or (\ref{xDefHP}) we have 
$b^2 = 2/a \pm x$, so substituting for $a$ in the above gives
 \begin{align}
   \text{HX:} \quad t & = t_B + x^{-3/2}  \left\{
      \sqrt{ \left( \frac{b^2 + x}{b^2 - x} \right)^2 - 1}\;
      - \text{arcosh} \left( \frac{b^2 + x}{b^2 - x} \right) \right\} ;
      \label{t.of.Rt.Hyp.exp} \\
   \text{EX:} \quad t & = t_B + x^{-3/2}  \left\{
      \arccos \left( \frac{b^2 + x}{b^2 - x} \right)
      - \sqrt{ 1 - \left( \frac{b^2 + x}{b^2 - x} \right)^2}\; \right\} ;
      \label{t.of.Rt.Ell.exp} \\
   \text{EC:} \quad t & = t_B + x^{-3/2}  \left\{
      2 \pi - \arccos \left( \frac{b^2 + x}{b^2 - x} \right)
      + \sqrt{ 1 - \left( \frac{b^2 + x}{b^2 - x} \right)^2}\; \right\} ;
      \label{t.of.Rt.Ell.col} \\
   \text{HC:} \quad t & = t_C - x^{-3/2}  \left\{
      \sqrt{ \left( \frac{b^2 + x}{b^2 - x} \right)^2 - 1}\;
      - \text{arcosh} \left( \frac{b^2 + x}{b^2 - x} \right)
      \right\} .
      \label{t.of.Rt.Hyp.col}
 \end{align}
 \end{widetext}
There are also two borderline cases, the parabolic case, and the
elliptic case which has reached maximum expansion at time $t$. Since the
above expressions are not numerically well-behaved near these
borderlines, we will present series solutions for near-parabolic (nP),
and for near-maximum-expansion (nMX) models.

The time reversed solutions, in which the hyperbolic and parabolic
solutions are collapsing (HC \& nPC), are also possible, but not as
relevant for cosmology.  Thus they will only be given for certain cases
to make a complete listing of possibilities.

It is convenient in what follows to use $M(r)$ as the radial coordinate
(i.e. $r = M(r)$), since we are not really considering vacuum, or
extrema in the spatial section (``bellies" and ``necks").  Thus we can
integrate (\ref{rhoLT}) along a constant time slice, $t = t_i$, to
obtain
 \begin{align}
   R_i^3 = a_i^3 M = \int_0^M \frac{6}{\kappa \rho_i(\widetilde{M})} \, \d
   \widetilde{M} .
   \label{rhoRM}
 \end{align}
This equation tells us that, if we have a density profile $\rho_i(M) >
0$ as a function of mass given at a particular time $t_i$, then a
straightforward integration gives us $R_i(M)$ on that time slice.  Ways
of coping with regions of zero density were discussed in
\cite{KrHe2003a}.

   Ideally, we seek LT models that have regular origins, and are free of
shell crossings and surface layers \cite{HeLa1985}.

   See \cite{Kras1997,Bond1947} as well as
\cite{KrHe2002,KrHe2003a,KrHe2003b} for more details on LT models.

 \subsection{Profile to Profile Solutions}

   As background to the present work, we very briefly summarise the
results of the previous papers.

   Paper I showed that, if  a spherically symmetric density profile is
given at two different times, i.e.
 \begin{align}
   \rho & = \rho_1(M) > 0 \quad \text{at} \quad t = t_1 , \\
   \rho & = \rho_2(M) > 0 \quad \text{at} \quad t = t_2 ,
 \end{align}
then there always exists a LT model that evolves from one to the other.
The formulas for the arbitrary functions $E(M)$ and $t_B(M)$ were given,
as well as the conditions that determine which type of evolution applies
at each $M$ value.  Since the formula for $E(M)$ can only be given
implicitly, a numerical code was written to implement this algorithm.
An example of the formation of an Abell cluster with a realistic density
profile at the present day, starting from a small fluctuation at
recombination, was calculated and its evolution plotted.

   Although the formulas given were for $t_2 > t_1$ and for an expanding
model at $t_1$, it is easy to adapt to the time-reverse scenario.  A key
step in the solution process is converting a given density profile
$\rho_i(M)$ to an areal radius profile $R_i(M)$ via equation
(\ref{rhoRM}).  Thus, if $R_i(M)$ were given instead, this is easily
incorporated into the method.

   Since the two density profiles are arbitrary, it is entirely possible
that the resulting LT model could develop shell crossings somewhere, but
it is easy to check for this once $E(M)$ and $t_B(M)$ are known.  A second
possibility to check for is whether $2E$ has reached $-1$, which indicates
a maximum in the spatial sections, $R_{,r} = 0$ has been reached.
Assuming both given densities are finite and non-zero at this point, it
would be a regular comoving maximum, $R_{,r}(t,M_{\text{max}}) = 0$ for all
$t$.  Beyond this locus, a regular model would have $M$ and $R$ decreasing
with increasing radial distance, so, in order to make further progress,
one would have to replace $\rho(M)$ with $\rho(2 M_{\text{max}} - M)$.
But this eventuality was not included in the numerics, as models of
objects that large were not contemplated.

   Paper II considered the possibility that velocity profiles might be
given,
 \begin{align}
   {R_{,t}} & = (R_{,t})_i(M) > 0 \quad \text{at} \quad t = t_i ,
 \end{align}
 and again showed how to find the LT model that evolves between two such
profiles or between one density and one velocity profile.  Several numerical
examples were given, illustrating different possibilities, including an improved
Abell cluster model and others that showed how radical changes in the density
profile are possible, and highlight the fact that both density and velocity
fluctuations at recombination play a significant role.  There was also a model
of the development of a void, but the density and velocity fluctuations at
recombination could not both be made small enough in any of the models tried.
Improved void models were constructed in \cite{BoKrHe2005}, using a realistic
 present-day density profile based on observations, and various density or velocity 
profiles at last scattering consistent with CMB observations, as well as a variety 
of model parameters, particularly $\Omega_{\text{matter}}$ and $\Omega_\Lambda$, 
$H_0$, $k$.  It was found that in each model there was always some inconsistency 
with observations: either the void density was not low enough, or the initial 
velocity was too high, or the density profile at the void wall was too steep and 
developed a shell crossing.  A much improved consistency with observations was 
achieved by Bolejko \cite{Bole2005} by including an inhomogeneous radiation component 
in the density distribution after last scattering.

   Paper III used the above methods to generate a model of a galaxy with a
central black hole evolving from a small fluctuation at recombination.
The final density profile was made a good fit to present day
observations with data from M87, whose central black hole could be as
massive as $3 \times 10^9~M_\odot$.  The initial fluctuations at
recombination were well
within observational limits%
 \footnote{The amplitudes of the density and of the velocity perturbation
were within the limits set by observations done at the scale of $\approx
1^{\circ}$.  However, the scale appropriate for a single galaxy is
$\approx 0.004^{\circ}$, and at this scale there are no observational data
at all.}%
 .  Two possibilities were considered for the central black hole.  The
first supposed that it formed by collapse of matter during the evolution of the
model, and for the particular model chosen, we found the black hole formed about
$4 \times 10^8$ years before the present.  The second supposed that it was a
full Schwarzschild-Kruskal-Szekeres type wormhole with past and future
singularities, but filled with matter \cite{Hell1987}.  Again for the particular
model chosen, it was found that the wormhole is only open for $6 \times
10^{-5}$~s, and the original black hole mass was only $2~M_\odot$, with all the
rest of the mass accreting onto it over time.  By recombination the mass was $2
\times 10^5~M_\odot$, with a horizon $5 \times 10^{-3}$~AU across, which
corresponds to $4 \times 10^{-13}$ degrees on the CMB sky --- much too small to
leave an observable imprint (see Paper II for formulae relating the sizes of
various objects to the angles that their images fill in the CMB sky).

   The subsequent sections complement these results by adding other
criteria by which models can be specified, and deriving the solution
algorithms that give the LT arbitrary functions in each case.

 \section{Models with a Simultaneous Bang Time}

We here show how to find a LT model that evolves to a given `final' (or
`initial' or `middle') density profile starting from a simultaneous bang
time
 \begin{align}
   t_B' = 0 , \qquad (t_B = \text{constant}) .
 \end{align}
 By (\ref{Rtsq}), the value of $E$ is unimportant near $R = 0$, so the
bang is Robertson-Walker (RW) like.  This condition is known to generate only
growing modes, but as discussed below, does not quite cover all possibilities,
though the omitted case is not relevant to cosmology in an expanding universe.

 \subsection{Density Profile Given at Time $t_i$}

We specify a density profile $\rho_i(M)$ at time $t_i$, and choose
$t_B$.  The function $a_i(M)$ is then determined from $\rho_i(M)$ via
(\ref{rhoRM}) and (\ref{aiDef}), and the equations to be solved, in the
three cases, are (\ref{t.of.R.Ell.exp}), (\ref{t.of.R.Ell.col}) \&
(\ref{t.of.R.Hyp.exp}) with $t_B =$ constant:
 \begin{align}
   \text{HX:} && 0 & = \sqrt{(1 + a_i x)^2 - 1}\;
      - \text{arcosh}(1 + a_i x) \nn \\
   &&&~~~~ - x^{3/2} (t_i - t_B) \df \psi_{BDH}(x) ;
      \label{SimBangDenHeq} \\
   \text{EX:} && 0 & = \arccos(1 - a_i x)
      - \sqrt{1 - (1 - a_i x)^2}\; \nn \\
   &&&~~~~ - x^{3/2} (t_i - t_B) \df \psi_{BDX}(x) ;
      \label{SimBangDenEXeq} \\
   \text{EC:} && 0 & = 2 \pi - \arccos(1 - a_i x)
      + \sqrt{1 - (1 - a_i x)^2}\; \nn \\
   &&&~~~~ - x^{3/2} (t_i - t_B) \df \psi_{BDC}(x) .
      \label{SimBangDenECeq}
 \end{align}
 Here the age of the model, $(t_i - t_B)$ is a free parameter that must
be specified along with $a_i$.  Equations
 (\ref{SimBangDenEXeq})-(\ref{SimBangDenHeq}) may be solved numerically
for $x(M)$ and hence $E(M)$.  In addition, we give series expansions for
solutions that are near to an expanding parabolic model (nPX), and near
to maximum expansion in an elliptic model (nEM) at $t_i$, because the above
solutions would encounter numerical difficulties close to the parabolic
and maximum expansion borderlines:
 \begin{align}
   \text{nPX:} && x & \approx \frac{20}{3 a_i} \left(
      - \frac{d\tau_P}{\tau_P} + \frac{25 d\tau_P^2}{14 \tau_P^2} \right) ;
      \label{SimBangDenPeq} \\
   \text{nEM:} && x & \approx \frac{2}{a_i} - \frac{d\tau_N^2}{a_i^4}
      + \frac{3 \pi d\tau_N^3}{2^{5/2} a_i^{11/2}} ;
      \label{SimBangDenEMeq}
 \end{align}
 where
 \begin{align}
   \tau_i & = t_i - t_B ,
      \label{taui-def} \\
   \tau_P & = \frac{\sqrt{2}\;}{3} a_i^{3/2} ,
      \label{tauP-def} \\
   d\tau_P & = \tau_i - \tau_P
      \label{dtauP-def} , \\
   \tau_N & = \pi \left( \frac{a_i}{2} \right)^{3/2} ,
      \label{tauN-def} \\
   d\tau_N & = \tau_i - \tau_N .
      \label{dtauN-def}
 \end{align}
 One may think of $\tau_P$ as the time it would take a parabolic model to
reach $a_i$, and $\tau_N$ as the time to maximum expansion \underbar{if}
$a_i$ were the maximum $a$ value.  But note that $d\tau_N$ is not the time
since maximum expansion, because when $d\tau_N \neq 0$, the model is not
exactly at maximum expansion, so $a_i$ and $\tau_N$ are less than their
maximum values.  Note too that (\ref{xDefHP}) is used to define $x$ for
the near parabolic series, so a negative $x$ indicates a slightly elliptic
model.

 \subsubsection{Existence of Solutions}

   We next consider existence for each $M$ value point by point, as the
solution type does not have to be the same at each point.  Since the argument
is very similar to that of previous papers, we will only here consider one
case, and otherwise just summarise the conditions.

   We take $(t_i - t_B) > 0$, and, for the EX case (\ref{SimBangDenEXeq}), 
we calculate
 \begin{gather}
   \psi_{BDX}(0) = 0 , \nn \\
   \lim_{x \to 2/a_i} \psi_{BDX} = \pi - (2/a_i)^{2/3} (t_i - t_B) ,
      \label{psiGDX.at.x=2/a} \\
   \td{\psi_{BDX}}{x} = \psi_{BDX,x} =
      \sqrt{x}\; \left[ \frac{a_i^{3/2}}{\sqrt{2 - a_i x}\;}
      - \frac{3}{2} (t_i - t_B) \right] , \nn \\
   \psi_{BDX,x}(0) = 0 , \nn \\
   \lim_{x \to 2/a} \psi_{BDX,x} = \infty . \nn
 \end{gather}
Clearly $\psi_{BDX,x}$ is the product of a positive term ($\sqrt{x}\;$)
and a monotonically increasing term (in square brackets), and therefore
it can only change sign if the second term is negative at $x = 0$, i.e.
 \begin{align}
   \frac{\sqrt{2}\; a_i^{3/2}}{3} < (t_i - t_B) .
   \label{BDXvalidLow}
 \end{align}
Only in this case does $\psi_{BDX}$ descend below $0$ before rising to
the limit (\ref{psiGDX.at.x=2/a}), and only if this limit is positive,
i.e. if
 \begin{align}
   (t_i - t_B) < \pi \left( \frac{a_i}{2} \right)^{3/2},
   \label{BDXvalidHigh}
 \end{align}
 does it create a root at $x > 0$.  Otherwise $\psi_{BDX}$ either rises
monotonically from the intial $0$ at $x = 0$, or it descends from $0$ and
then rises to a negative limiting value.  Therefore (\ref{BDXvalidLow})
and (\ref{BDXvalidHigh}) are the necessary and sufficient conditions for a
growing mode, 
 still-expanding, elliptic solution to exist.  The complete set of
conditions is
 \begin{align}
   \text{HX:} \quad & (t_i - t_B) < \frac{\sqrt{2}\; a_i^{3/2}}{3} ; \\
   \text{PX:} \quad & (t_i - t_B) = \frac{\sqrt{2}\; a_i^{3/2}}{3} ; \\
   \text{EX:} \quad & \frac{\sqrt{2}\; a_i^{3/2}}{3} < (t_i - t_B)
      < \pi \left( \frac{a}{2} \right)^{3/2} ; \\
   \text{EM:} \quad & (t_i - t_B) = \pi \left( \frac{a}{2} \right)^{3/2} ; \\
   \text{EC:} \quad & \pi \left( \frac{a}{2} \right)^{3/2} < (t_i - t_B) .
 \end{align}
 The parabolic and maximum expansion borderlines are sufficiently obvious
that they will not be listed in subsequent sets of existence conditions.

 \subsection{Velocity Profile Given at Time $t_i$}

   For this scenario, the equations to be solved are
(\ref{t.of.Rt.Ell.exp}), (\ref{t.of.Rt.Ell.col}) \& (\ref{t.of.Rt.Hyp.exp})
with $t_B =$ constant:
 \begin{align}
   \text{HX:} && 0 & =
      \sqrt{ \left( \frac{b_i^2 + x}{b_i^2 - x} \right)^2 - 1 }\;
      - \text{arcosh} \left( \frac{b_i^2 + x}{b_i^2 - x} \right) \nn \\
   &&&~~~~ - x^{3/2} (t_i - t_B) \df \psi_{BVH}(x) ;
      \label{SimBangVelHeq} \\
   \text{EX:} && 0 & =
      + \arccos \left( \frac{b_i^2 - x}{b_i^2 + x} \right)
      - \sqrt{ 1 - \left( \frac{b_i^2 - x}{b_i^2 + x} \right)^2 }\; \nn \\
   &&&~~~~ - x^{3/2} (t_i - t_B) \df \psi_{BVX}(x) ;
      \label{SimBangVelEXeq} \\
   \text{EC:} && 0 & =
      2 \pi - \arccos \left( \frac{b_i^2 - x}{b_i^2 + x} \right)
      + \sqrt{ 1 - \left( \frac{b_i^2 - x}{b_i^2 + x} \right)^2 }\; \nn \\
   &&&~~~~ - x^{3/2} (t_i - t_B) \df \psi_{BVC}(x) ;
      \label{SimBangVelECeq}
 \end{align}
 where $b_i = (R_{,t})_i/M^{1/3}$, which are to be solved numerically, and, 
from series expansions near the borderlines:
 \begin{align}
   \text{nPX:} && x & \approx \frac{5 b_i^2}{6} \left(
      \frac{d\tau_P}{\tau_P} - \frac{25 d\tau_P^2}{28 \tau_P^2} \right) ;
      \label{SimBangVelPeq} \\
   \text{nEM:} && x & \approx x_N \left( 1
       + \frac{16}{3 \pi^2} \frac{b_i}{\overline{b}}
       + \frac{64}{3 \pi^4} \frac{b_i^2}{\overline{b}^2} \right) ;
      \label{SimBangVelEMeq}
 \intertext{where}
   &&&~~~~ \tau_i = t_i - t_B , \\
   &&&~~~~ \tau_P = \frac{4}{3 b_i^3} ,
      \label{tauP-b-def} \\
   &&&~~~~ d\tau_P = \tau_i - \tau_P , \\
   &&&~~~~ x_N = \left( \frac{\pi}{\tau_i} \right)^{2/3} ,
      \label{xN-def} \\
   &&&~~~~ \overline{b} = \frac{2 \sqrt{x_N}\;}{\pi} .
      \label{b-bar-def}
 \end{align}
 Here $\overline{b}$ is a kind of average velocity; if the model were exactly
at maximum expansion after $\tau_i$, then $a$ would be $a_N = 2/x_N$ and
$\overline{b} = \frac{a_N}{\tau_i}$.

 \subsubsection{Existence of Solutions}

   The existence conditions for each case are:
 \begin{align}
   \text{HX:} \quad & b_i > 0 , \quad (t_i - t_B) > \frac{4}{3 b_i^3} ; \\
   \text{EX:} \quad & b_i > 0 , \quad (t_i - t_B) < \frac{4}{3 b_i^3} ; \\
   \text{EC:} \quad & b_i < 0 .
 \end{align}

 \section{Models that Become Homogeneous at Late Times}

   We next find a LT model that evolves from a given `initial' profile and
approaches a RW model at late times, $t \to \infty$.  Only expanding
hyperbolic (and parabolic) models have an infinite future, and for these
we require
 \begin{align}
   2 E = K \, M^{2/3} ~,~~~~~~~~ K~\mbox{a constant} .
   \label{LateRWConditH}
 \end{align}
 Clearly, this case has only decaying modes.  The time reverse of this case
 --- the collapsing hyperbolic model
 --- also satisfies the same condition but was RW like in the infinite past.  
See \cite{Kras1997} for examples.

This condition may also be applied to elliptic models, and although the
inhomogeneities due to $t_{B,r} \neq 0$ do decay initially, other modes
can grow as the big crunch is approached because $t_{C,r} \neq 0$.
However, (\ref{LateRWConditH}) has the effect of making the lifetime
$T(r)$ along the dust worldlines a constant (see (\ref{Lifetime})). Now,
it is known that the set of necessary and sufficient conditions for no
shell crossings in an elliptic model is $\{t_{B,r}/M_{,r} \leq 0,
t_{C,r}/M_{,r} \geq 0\}$ \cite{HeLa1985}, but $T(r) =$ const makes it
impossible to obey both these inequalities, unless $t_B =$ const and
$t_C =$ const, in which case the Friedmann model results.  Thus the only
elliptic model without shell crossings that satisfies
(\ref{LateRWConditH}) is the RW model%
 \footnote{
 Nevertheless, part of the evolution of
 non-homogeneous elliptic models obeying (\ref{LateRWConditH}) will be free 
of shell crossings and may be of interest.}%
 .

 \subsection{Density Profile Given at Time $t_i$}

Applying (\ref{LateRWConditH}) to (\ref{t.of.R.Hyp.exp}) leads to the direct
solution
 \begin{align}
   \text{HX:} \quad && t_B & = t_i - \big[ \sqrt{(1 + a_i K)^2 - 1}\; \nn \\
   &&&~~~~ - \text{arcosh}(1 + a_i K) \big] \big/ K^{3/2} ;
      \label{LateRWDenHeq} \\
   \text{nPX:} \quad && t_B & \approx t_i - \sqrt{2 a_i^3} \left(
      \frac{1}{3} - \frac{a_i K}{20} + \frac{3 a_i^2 K^2}{224} \right) .
      \label{LateRWDenPeq}
 \end{align}
 Notice that the model is only fully defined once $K$ is given
 --- in other words, the limiting late time RW model must be fully
specified.

 \subsubsection{Existence of Solutions}

The condition is
 \begin{align}
   \text{HX:} \quad & K > 0 ,
      \label{LateRWDenSolExistH}
 \end{align}
 and it follows that
 \begin{align}
   t_i - t_B < \frac{\sqrt{2}}{3} a_i^{3/2}.
      \label{LateRWDenSolConsequenceH}
 \end{align}
 This will keep appearing in what follows as a complementary inequality
to the others we shall derive.

 \subsection{Velocity Profile Given at Time $t_i$}

   Given the velocity distribution $b_i(M)$, the hyperbolic case again has
a direct solution for $t_B$, once (\ref{LateRWConditH}) is applied,
 \begin{align}
   \text{HX:} \quad && t_B & = t_i - \Bigg[ \sqrt{
         \left( \frac{b_i^2 + K}{b_i^2 - K} \right)^2 - 1 }\; \nn \\
      &&&~~~~ - \text{arcosh} \left( \frac{b_i^2 + K}{b_i^2 - K}
         \right) \Bigg] \Bigg/ K^{3/2} ;
      \label{LateRWVelHeq} \\
   \text{nPX:} \quad && t_B & \approx t_i - \frac{4}{b_i^3} \left(
      \frac{1}{3} + \frac{2 K}{5 b_i^2} + \frac{3 K^2}{7 b_i^4} \right) .
      \label{LateRWVelPeq}
 \end{align}

 \subsubsection{Existence of Solutions}

 \begin{align}
   \text{HX:} \quad & b_i > 0 , \quad 0 < K < b_i^2, \quad (t_i - t_B) >
      \frac{4}{3 b_i^3}.
      \label{LateRWVelSolExistH}
 \end{align}

 \section{Models with a simultaneous Crunch Time}

   For elliptic models, the requirement of only decaying modes is that the 
crunch time must be simultaneous (i.e. the crunch be RW like),
 \begin{align}
   t_C = \text{constant} = t_B + \frac{2 \pi M}{(-2E)^{3/2}}
   = t_B + \frac{2 \pi}{x^{3/2}} .
   \label{SimCrunch}
 \end{align}
 H \& P models with a simultaneous crunch time obviously have no bang, and 
are therefore collapsing at all times.

 \subsection{Density Profile Given at Time $t_i$}

 The equations to be solved are (\ref{t.of.R.Ell.exp}) \&
(\ref{t.of.R.Ell.col}) with (\ref{SimCrunch}) and $t_C =$ constant:
 \begin{align}
   \text{EX:} && 0 & = - 2 \pi + \arccos(1 - a_i x)
      - \sqrt{1 - (1 - a_i x)^2}\; \nn \\
      &&&~~~~ + x^{3/2} (t_C - t_i)
      \df \psi_{CDX}(x) ;
      \label{SimCrunchDenEXeq} \\
   \text{EC:} && 0 & = - \arccos(1 - a_i x)
      + \sqrt{1 - (1 - a_i x)^2}\; \nn \\
      &&&~~~~ + x^{3/2} (t_C - t_i)
      \df \psi_{CDC}(x) ;
      \label{SimCrunchDenECeq} \\
   \text{HC:} && 0 & = \sqrt{(1 + a_i x)^2 - 1}\;
      - \text{arcosh}(1 + a_i x) \nn \\
   &&&~~~~ - x^{3/2} (t_C - t_i)
      \df \psi_{CDH}(x) ;
      \label{SimCrunchDenHCeq}
 \intertext{with the equations for the borderline cases being:}
   \text{nPC:} && x & \approx \frac{20}{3 a_i} \left(
      - \frac{d\tau_P}{\tau_P} + \frac{25 d\tau_P^2}{14 \tau_P^2} \right) ;
      \label{SimCrunchDenPCeq} \\
   \text{nEM:} && x & \approx \frac{2}{a_i} - \frac{d\tau_N^2}{a_i^4}
      + \frac{3 \pi d\tau_N^3}{2^{5/2} a_i^{11/2}} ;
      \label{SimCrunchDenEMeq}
 \end{align}
 where
 \begin{align}
   \tau_i & = t_C - t_i ,
      \label{tauf-def}
 \end{align}
 while (\ref{tauP-def})--(\ref{dtauN-def}) define $\tau_P$, $\tau_N$ and 
$d\tau_N$.  Again the remaining lifetime of the model, $(t_C - t_i)$ is a 
free parameter that must be specified along with $a_i$, and as before, eqs.
(\ref{SimCrunchDenEXeq})--(\ref{SimCrunchDenHCeq}) are to be solved 
numerically for $x$.  The functions $E$ and $t_B$ then follow from 
(\ref{xDefHP}) or (\ref{xDefE}) and (\ref{SimCrunch}).

 \subsubsection{Existence of Solutions}

   The complete set of conditions is
 \begin{align}
   \text{EX:} \quad & \pi \left( \frac{a_i}{2} \right)^{3/2} <
      (t_C - t_i) ; \\
   \text{EC:} \quad & \frac {\sqrt{2}} 3\ {a_i}^{3/2} <
      (t_C - t_i) < \pi \left( \frac{a_i}{2} \right)^{3/2} ;
      \label{SimCrunchDenSolExistEC} \\
   \text{HC:} \quad & (t_C - t_i) < \frac {\sqrt{2}} 3\ {a_i}^{3/2}.
      \label{SimCrunchDenSolExistH}
 \end{align}
 The first inequality in (\ref{SimCrunchDenSolExistEC}) means that the
time-difference between $t_i$ and $t_C$ is larger than it would be in a
collapsing parabolic model --- this is seen from (\ref{ParEv}).  If
$(t_C - t_i)$ is smaller, then the model collapsing from the given
density distribution to the Big Crunch in the time-interval $(t_C -
t_i)$ would have to be hyperbolic.  This is consistent with the
time-reverse of Eq. (\ref{LateRWDenSolConsequenceH}).

 \subsection{Velocity Profile Given at Time $t_i$}

 These models are found by solving
 \begin{align}
   \text{EX:} && 0 & =  - 2 \pi
      + \arccos \left( \frac{b_i^2 - x}{b_i^2 + x} \right)
      - \sqrt{ 1 - \left( \frac{b_i^2 - x}{b_i^2 + x} \right)^2 }\; \nn \\
   &&&~~~~ + x^{3/2} (t_C - t_i)
      \df \psi_{CVX}(x) ;
      \label{SimCrunchVelEXeq} \\
   \text{EC:} && 0 & =
      - \arccos \left( \frac{b_i^2 - x}{b_i^2 + x} \right)
      + \sqrt{ 1 - \left( \frac{b_i^2 - x}{b_i^2 + x} \right)^2 }\; \nn \\
   &&&~~~~ + x^{3/2} (t_C - t_i).
      \df \psi_{CVC}(x) ; \\
   \text{HC:} && 0 & = \sqrt{ \left( \frac{b_i^2 + x}{b_i^2 - x}
   \right)^2  - 1}\; - \text{arcosh} \left( \frac{b_i^2 + x}{b_i^2 - x}
   \right) \nn \\
   &&&~~~~ - x^{3/2} (t_C - t_i)
      \df \psi_{CVH}(x) ;
 \intertext{or, for the borderline cases, by calculating}
   \text{nPX:} && x & \approx \frac{5 b_i^2}{6} \left(
      \frac{d\tau_P}{\tau_P} - \frac{25 d\tau_P^2}{28 \tau_P^2} \right) ;
      \label{SimCrunchVelPeq} \\
   \text{nEM:} && x & \approx x_N \left( 1
       + \frac{16}{3 \pi^2} \frac{b_i}{\overline{b}}
       + \frac{64}{3 \pi^4} \frac{b_i^2}{\overline{b}^2} \right) ;
      \label{SimCrunchVelEMeq}
 \intertext{where}
   &&&~~~~ \tau_i = t_C - t_i ,
 \end{align}
 while (\ref{tauP-b-def}), (\ref{xN-def}) and (\ref{b-bar-def}) define
$\tau_P$, $x_N$ and $\overline{b}$.

 \subsubsection{Existence of Solutions}

 \begin{align}
   \text{EX:} \quad & b_i > 0 , \quad t_C > t_i ;
      \label{SimCrunchVelSolExistEX} \\
   \text{EC:} \quad & b_i < 0 , \quad (t_C - t_i) < \frac{4}{3 b_i^3} ;
      \label{SimCrunchVelSolExistEC} \\
   \text{HC:} \quad & b_i < 0 , \quad (t_C - t_i) > \frac{4}{3 b_i^3} ;
      \label{SimCrunchVelSolExistHC}
 \end{align}
where the last equation above is for collapsing hyperbolic models.  By
(\ref{SimCrunch}), $x^{3/2} (t_C - t_i) = 2 \pi - x^{3/2} (t_i - t_B)$,
and using this in (\ref{SimCrunchVelEXeq}) leads to an equivalent
existence condition for the EX case, $\quad b_i > 0 , \quad (t_i - t_B)
> \frac{4}{3 b_i^3} \quad$.  This means the conditions
(\ref{LateRWVelSolExistH}) and (\ref{SimCrunchVelSolExistEX}) are
mutually exclusive.

 \section{Growing and Decaying Modes}

   For pure decaying modes, hyperbolic (and parabolic) models that are
expanding must become RW like at late times, whereas for elliptic models
and for hyperbolic and parabolic models that are collapsing, they must
become RW like at the crunch.

   Conversely, for pure growing modes, hyperbolic models that are
collapsing must have been RW like in the distant past, while elliptic models and
expanding hyperbolic and parabolic models must have been RW like at the bang.

   This is summarised in the following table: \\
 \centerline{\begin{tabular}{l|l|l}
 & only & only \\
 & growing & decaying \\
 & modes & modes \\
 \hline
 HX, PX & $t_{B,r} = 0$     & $2 E = K M^{2/3}$ \\
 E      & $t_{B,r} = 0$     & $t_{C,r} = 0$ \\
 PC, HC & $2 E = K M^{2/3}$ & $t_{C,r} = 0$ \\
 \end{tabular}}

 \section{Models with a Simultaneous Time of Maximum Expansion}

   Not infrequently, authors seeking a manifestly regular initial condition
for an inhomogeneous matter distribution, have required a finite density
distribution and zero initial velocity.  This is achieved in an LT model
if the moment of maximum expansion occurs at the same time along all worldlines.

   Naturally only elliptic (EX \& EC) models have a moment of maximum expansion.
In general, the time of maximum expansion (\ref{t.MX}) is different for
each worldline.  The condition that it be simultaneous is
$t_{MX}=$~constant, i.e.
 \begin{align}
   t_{SMX} = \text{constant} = t_B + \frac{\pi M}{(-2E)^{3/2}} =
   t_B + \frac{\pi}{x^{3/2}} .
   \label{tB.SMX}
 \end{align}
 This needs to be combined with another condition to obtain a solution.

 \subsection{Density Profile Given at Time $t_i$}
 \label{SMX-DenProf}

   The equations to be solved, with $t_{SMX} =$ constant, are
 \begin{align}
   \text{EX:} && 0 & = - \pi + \arccos(1 - a_i x)
      - \sqrt{1 - (1 - a_i x)^2}\; \nn \\
      &&&~~~~ + x^{3/2} (t_{SMX} - t_i)
      \df \psi_{SDX}(x) ; \\
   \text{EC:} && 0 & = \pi - \arccos(1 - a_i x)
       + \sqrt{1 - (1 - a_i x)^2}\; \nn \\
      &&&~~~~ + x^{3/2} (t_{SMX} - t_i)
      \df \psi_{SDC}(x) .
 \end{align}
 As above, these equations are solved numerically for $x$, and $E$ and
$t_B$ then follow from (\ref{xDefE}) and (\ref{tB.SMX}).

 \subsubsection{Existence of Solutions}

   The complete set of conditions is
 \begin{align}
   \text{EX:} \quad & t_{SMX} > t_i ; \\
   \text{EC:} \quad & t_{SMX} < t_i .
 \end{align}

 \subsection{Density Profile Given at a Simultaneous Time of Maximum
Exapnsion}

   Although this is a special case of section \ref{SMX-DenProf}, it has an
explicit solution.  Given $\rho = \rho_{SMX}(M)$ at $t = t_{SMX}$, we
calculate $R_{SMX}(M)$ with (\ref{rhoRM}), then write (\ref{R.MX}) as
 \begin{align}
   (- E) = \frac{M}{R_{SMX}} ~~~~\Leftrightarrow~~~~
   x = \frac{2}{a_{SMX}}
 \end{align}
 and use it in (\ref{tB.SMX}), giving the direct solution
 \begin{align}
   t_B & = t_{SMX} - \pi \sqrt{\frac{a_{SMX}^3}{8}}\; , \\
   E & = - \frac{M^{2/3}}{a_{SMX}} .
 \end{align}
 Solutions obviously exist for $R_{SMX} > 0$ and $M > 0$.

 \subsection{Velocity Profile Given at Time $t_i$}

   The two elliptic cases are found from,
 \begin{align}
   \text{EX:} && 0 & =  - \pi
      + \arccos \left( \frac{b_i^2 - x}{b_i^2 + x} \right)
      - \sqrt{ 1 - \left( \frac{b_i^2 - x}{b_i^2 + x} \right)^2 }\; \nn \\
   &&&~~~~ + x^{3/2} (t_{SMX} - t_i)
      \df \psi_{SVX}(x) ; \\
   \text{EC:} && 0 & = \pi
      - \arccos \left( \frac{b_i^2 - x}{b_i^2 + x} \right)
      + \sqrt{ 1 - \left( \frac{b_i^2 - x}{b_i^2 + x} \right)^2 }\; \nn \\
   &&&~~~~ + x^{3/2} (t_{SMX} - t_i)
      \df \psi_{SVC}(x) .
 \end{align}

 \subsubsection{Existence of Solutions}

   The complete set of conditions is
 \begin{align}
   \text{EX:} \quad & b_i > 0 , \quad t_{SMX} > t_i ; \\
   \text{EC:} \quad & b_i < 0 , \quad t_{SMX} < t_i .
 \end{align}
 The conditions $b_i > 0$ and $b_i < 0$ follow directly from the
assumptions that the model is expanding or collapsing, respectively. The
other ones are conditions for solvability of the corresponding
equations.

 \section{Models with Given Density and Velocity profiles at the Same Time}

 For this case, we are given $R_{,t}(M, t_i)$ and $\rho(M, t_i)$ which provide
us with $b_i(M)$ and $a_i(M)$ via equations (\ref{rhoRM}), (\ref{aiDef}) and
(\ref{biDef}).  We then solve (\ref{Rtsq}) for $E$,
 \begin{align}
   2 E = (R_{,t})_i^2 - \frac{2 M}{R_i}
   ~~~~\Leftrightarrow~~~~
   \pm x = \frac{2 E}{M^{2/3}} = b_i^2 - \frac{2}{a_i} ,
   \label{DVsameE.eq}
 \end{align}
and obtain $t_B$ from one of (\ref{t.of.R.Hyp.exp})-(\ref{t.of.R.Hyp.col}).
The equation for $t_B$ is sensitive to the sign of $b_i$ (i.e,
$R_{,t}$), even though the $E$ equation isn't. This method is given in
\cite{BoKrHe2005}.

   The simple solution (\ref{DVsameE.eq}) does also follow as the limit
$t_2 \to t_1$ from the considerations of Paper II, though merely substituting 
$t_2 = t_1$ leads to degenerate equations.  See Appendix A for a proof.

   The borderline cases need no special treatment, as there are no numerical
difficulties arising from being close to them.  The model is parabolic if
$b_i^2 = 2/a_i$, and is at maximum expansion (at $t_i$) if $b_i = 0$.

 \setcounter{subsubsection}{0} 
 \subsubsection{Existence of Solutions}

   Eq (\ref{DVsameE.eq}) always has a solution, but, for the purposes of
determining $t_B$, the various types of solution exist if
 \begin{align}
   \mbox{HX:} \quad & b_i^2 > \frac{2}{a_i} ~~\mbox{and}~~ b_i > 0 ; \label{DVsameT.Hcond} \\
   \mbox{EX:} \quad & b_i^2 < \frac{2}{a_i} ~~\mbox{and}~~ b_i > 0 ; \label{DVsameT.EXcond} \\
   \mbox{EC:} \quad & b_i^2 < \frac{2}{a_i} ~~\mbox{and}~~ b_i < 0 ; \label{DVsameT.ECcond} \\
   \mbox{HC:} \quad & b_i^2 > \frac{2}{a_i} ~~\mbox{and}~~ b_i < 0 . \label{DVsameT.HCcond}
 \end{align}

 \section{Models with a Given Velocity Profile at Late Times}

   By ``late times'' we mean the asymptotic future, i.e. the limit $\eta
\to \infty$ and $t \to \infty$, so this section applies only to
expanding hyperbolic (HX) models.  The time reverse
 --- a collapsing hyperbolic (HC) model with a given density or velocity
profile in the infinite past
 --- follows in an obvious way.

   From (\ref{HypEv}) and (\ref{Rtsq}) we find
 \begin{align}
   \lim_{\eta \to \infty} \frac R {t - t_B} & = \sqrt{2E}\; ,
      \label{R.Hyp.Late} \\
   \lim_{t \to \infty} {{R_{,t}}} & = \sqrt{2E}\; ,
      \label{Rt.Hyp.Late}
 \end{align}
 which gives simply
 \begin{align}
   E = \frac{{R_{,t}}^2_{\text{late}}}{2}
   \Leftrightarrow
   x = b_{\text{late}}^2 .
      \label{xLateVel}
 \end{align}
This always exists, and fully determines $E(M)$, leaving $t_B(M)$ free
to be determined by a second requirement.

 \subsection{Density Profile Given at Time $t_i$}

   Since $x$ is known from (\ref{xLateVel}) and $t_i$, $\rho_i$ and $a_i$
are finte, we find $t_B$ from:
 \begin{align}
   \text{HX:} \quad t_B & = t_i - \big[
         \sqrt{(1 + a_i b_{\text{late}}^2)^2 - 1}\; \nn \\
      &~~~~~~~~~~ - \text{arcosh}(1 + a_i b_{\text{late}}^2) \big]
         \big/ b_{\text{late}}^3 ;
      \label{LateVelDenSolH} \\
   \text{nPX:} \quad t_B & \approx t_i - \tau_P \left( 1
      - \frac{3 a_i b_{\text{late}}^2}{20}
      + \frac{9 a_i^2 b_{\text{late}}^4}{224} \right)
      \label{LateVelDenSolP} ;
 \end{align}
 which are well defined for all $b_{\text{late}} \geq 0$, but imply
(\ref{LateRWDenSolConsequenceH}).

 \subsubsection{Existence of Solutions}

 The model is expanding if
 \begin{align}
   b_{\text{late}} > 0 ,
 \end{align}
 and the above solution is well defined provided
 \begin{align}
   a_i > 0 .
 \end{align}

 \subsection{Velocity Profile Given at Time $t_i$}

The equation for $t_B$ is
 \begin{align}
   \text{HX:} \quad t_B & = t_i - \Bigg[ \sqrt{ \left( \frac{b_i^2
         + b_{\text{late}}^2}{b_i^2 - b_{\text{late}}^2} \right)^2 - 1 }\;
         \nn \\
      &~~~~~~~~~~ - \text{arcosh} \left( \frac{b_i^2
         + b_{\text{late}}^2}{b_i^2 - b_{\text{late}}^2}
         \right) \Bigg] \Bigg/ b_{\text{late}}^3 ;
      \label{LateVelVelSolH} \\
   \text{nPX:} \quad t_B & \approx t_i - \tau_P \left( 1
      + \frac{6 b_{\text{late}}^2}{5 b_i^2}
      + \frac{9 b_{\text{late}}^4}{7 b_i^4} \right) .
      \label{LateVelVelSolP}
 \end{align}

 \subsubsection{Existence of Solutions}

 This solution is well defined (for an expanding model) provided
 \begin{align}
   b_i > b_{\text{late}} > 0 .
 \end{align}
 This does mean that not all choices of $b_{\text{late}}(M)$ and $b_i(M)$
are possible, but it should be easy to choose a suitable $b_i$ after
$b_{\text{late}}$ is fixed.

 \subsection{Simultaneous Bang Time Specified}

   As soon as the (constant) value of $t_B$ is given, we have all the
LT arbitrary functions $E(M)$ and $t_B(M)$.

 \subsubsection{Existence of Solutions}

 Existence only requires that the model be expanding, 
 \begin{align}
   b_{\text{late}} > 0 .
 \end{align}

 \section{Models with a Given Density Profile at Late Times}

   From (\ref{HypEv}) and (\ref{rhoLT}) we obtain, using $\p R/\p M =
R_{,r}/M_{,r}$,
 \begin{align}
   \frac{\pdil{R}{M}}{(t - t_B)} & = \sqrt{2 E}\; \Bigg[
      \left( \frac{1}{M} - \frac{dE/dM}{E} \right)
      \frac{(\cosh \eta - 1)}{(\sinh \eta - \eta)} \nn \\
   &~~~~ + \left( \frac{3 \, dE/dM}{2 E} - \frac{1}{M} \right)
      \frac{\sinh \eta}{(\cosh \eta - 1)} \Bigg] \nn \\
   &~~~~ - \frac{(2 E)^2 \, dt_B/dM}{M}
      \frac{\sinh \eta}{(\cosh \eta - 1) (\sinh \eta - \eta)} \Bigg] ,
   \label{R'exact} \\
   \kappa \rho (t - t_B)^3 & = \frac{2}{
      \left( \frac{R}{(t - t_B)} \right)^2
      \left( \frac{\pdil{R}{M}}{(t - t_B)} \right)} .
 \end{align}
 Assuming $dt_B/dM$ is finite%
 \footnote{
 One can consider models in which $dt_B/dM$ becomes infinite while $dE/dM$
doesn't, either asymptotically or at individual points. However, at such
locations there is a permanent zero in the density, and the above limit
is not valid. The other cases that give a permanent zero in the density,
loci where $dE/dM$ is divergent and $dt_B/dM$ isn't, and loci where
$M_{,r} = 0$ are not problematic for this limit. (See Paper II for a
discussion of regions of zero density.)
 }%
 , we find that
 \begin{align}
   \lim_{\eta \to \infty}\ \frac {\pdil{R}{M}}{(t - t_B)}
   & = \frac{dE/dM}{\sqrt{2E}\;},
   \label{Rr.Hyp.Late}
 \end{align}
 which with (\ref{R.Hyp.Late}) leads to
 \begin{align}
\lim_{\eta \to \infty} \left[\kappa \rho (t - t_B)^3\right] = \frac{2}
{\sqrt{-2E}\; dE/dM}.
 \end{align}
Clearly $\rho (t - t_B)^3$ freezes out --- becomes time independent ---
and
 \begin{align}
   \lim_{\eta \to \infty} \kappa \rho (t - t_B)^3
   = \frac{6}{\td{}{M} \left( (2E)^{3/2}) \right)}.
 \end{align}
 Thus, if we specify the late time limit $[\rho (t -
t_B)^3]_{\text{late}}(M)$, we have
 \begin{align}
   x_{\text{late}}^{3/2} = \frac{(2E)^{3/2}}{M} = \frac{1}{M}
   \int_0^M \frac{6}{\kappa [\rho (t - t_B)^3]_{\text{late}}} \, \text{d}
   \widetilde{M} .
   \label{LateTimeHypDensitySoln}
 \end{align}
 This assumes an origin exists, i.e. $E = 0$ at $M = 0$.  If not, then we
must specify some $E = E_i$ at some $M = M_i$.

Again, (\ref{LateTimeHypDensitySoln}) fully determines $E(M)$, leaving
$t_B(M)$ free.

\subsection{Density Profile Given at Time $t_i$}

   In this case we find $t_B$ from:
 \begin{align}
   \text{HX:} \quad t_B & = t_i - \big[
      \sqrt{(1 + a_i x_{\text{late}})^2 - 1}\; \nn \\
      &~~~~~~~~~~ - \text{arcosh}(1 + a_i x_{\text{late}}) \big]
         \big/ x_{\text{late}}^{3/2} ;
      \label{LateDenDenSolH} \\
   \text{nPX:} \quad t_B & \approx t_i - \tau_P \left( 1
      - \frac{3 a_i x_{\text{late}}}{20}
      + \frac{9 a_i^2 x_{\text{late}}^2}{224} \right) ;
      \label{LateDenDenSolP}
 \end{align}
 which is well defined for all $x_{\text{late}} > 0$,
but again implies (\ref{LateRWDenSolConsequenceH}).

 \subsubsection{Existence of Solutions}

   Equation (\ref{LateTimeHypDensitySoln}) will have a positive solution
for $x$ for any positive $[\rho (t - t_B)^3]_{\text{late}}(M)$ and the
above equation for $t_B$ is well defined for any $a_i$ that is derived from
a positive $\rho_i$.

 \subsection{Velocity Profile Given at Time $t_i$}

   The equation for $t_B$ is
 \begin{align}
   \text{HX:} \quad t_B & = t_i - \Bigg[ \sqrt{ \left( \frac{b_i^2
         + x_{\text{late}}}{b_i^2 - x_{\text{late}}} \right)^2 - 1 }\;
         \nn \\
      &~~~~~~~~~~ - \text{arcosh} \left( \frac{b_i^2
         + x_{\text{late}}}{b_i^2 - x_{\text{late}}}
         \right) \Bigg] \Bigg/ x_{\text{late}}^{3/2} ;
      \label{LateDenVelSolH} \\
   \text{nPX:} \quad t_B & \approx t_i - \tau_P \left( 1
      + \frac{6 x_{\text{late}}}{5 b_i^2}
      + \frac{9 x_{\text{late}}^2}{7 b_i^4} \right) .
      \label{LateDenVelSolP}
 \end{align}

 \subsubsection{Existence of Solutions}

 The above is well defined provided
 \begin{align}
   b_i^2 > x_{\text{late}} .
 \end{align}
 It should not be hard to choose $b_i$ to satisfy this, once
$x_{\text{late}}$ is known.

 \subsection{Simultaneous Bang Time Specified}

   Again, setting the constant value of $t_B$ completes the specification 
of the LT arbitrary functions, and the solution always exists.

 \subsection{Late-Time Velocity Profile Given}

   Since specifying the 
 late-time velocity profile via eq. (\ref{xLateVel}),  and specifying the 
 late-time density profile via eq. (\ref{LateTimeHypDensitySoln}) both fix 
$x$, this combination is not possible.  (Should it happen that the two 
specifications are consistent, then another `boundary condition' would be 
required to specify a particular LT model.)

 \section{Conclusions}

   We have developed several new ways of specifying the `boundary' data needed
to uniquely determine the evolution of a
 {\LT} model, and thus provided more options for designing models 
with particular properties or behaviours.  Thus one can now easily generate models 
that start from an initial stationary state, or have only growing modes, or 
approach a specified density or velocity profile in the asymptotic future, or 
approach RW models at late times or diverge from them, etc, as listed in the 
introduction.  The foregoing properties are combined in pairs to fully specify 
a particular LT model.  Although several of the individual properties considered 
here have previously been used, what is significant here is the equations that 
result from the many combinations of pairs of properties, and the derivation of 
existence conditions for the 3 types of solution.  Also, our results have been 
presented in a form that is easily converted into coding for numerical calculations.  
 Current work \cite{BolHel05} provides an example of the use of some of the new LT 
specification methods to create and evolve a model of the Shapley concentration 
and the great attractor.

 \begin{acknowledgments}
   The research of AK was supported by the Polish Research Committee grant
no 2 P03B 12 724.  CH thanks the South African National Research
Foundation for a grant.  An award from the Poland-South Africa Technical
Cooperation Agreement is gratefully acknowledged.
 \end{acknowledgments}

 \appendix

 \section{Derivation of (\ref{DVsameE.eq}) from Paper II Results}

Using the results of Sec. IV of Paper II with $t_2 = t_1 = t_i$, and
remembering that for the EC case {\it both} profiles must be specified
in the collapsing phase, we obtain
 \begin{align}
   &\mbox{HX:} \nn \\
   &0 = \sqrt{ \left( 1 + a_i x \right)^2 - 1}\;
         - \text{arcosh}(1 + a_i x) \nn \\
      &~~ - \sqrt{ \left( \frac{b_i^2 + x}{b_i^2 - x} \right)^2 - 1}\;
         + \text{arcosh} \left( \frac{b_i^2 + x}{b_i^2 - x} \right) ;
         \label{DenVelSameTimeHeq} \\
   &\mbox{EX and EC:} \nn \\
   &0 = \arccos(1 - a_i x) - \sqrt{1 - (1 - a_i x)^2}\;
         \nn \\
      &~~ + \sqrt{1 - \left( \frac{b_i^2 - x}{b_i^2 + x}
         \right)^2}\;
         - \arccos \left( \frac{b_i^2 - x}{b_i^2 + x} \right) .
      \label{DenVelSameTimeEXeq}
 \end{align}
Denote $1 + a_i x = \cosh u$ and $\frac {{b_i}^2 + x} {{b_i}^2 - x} =
\cosh v$ for the hyperbolic case. Equation (\ref{DenVelSameTimeHeq})
then is equivalent to $\sinh u - u = \sinh v - v$. But the function
$F(y) := \sinh y - y$ is single-valued, so the equation $F(u) = F(v)$
has only one solution, $u = v$. In our case, this solution is ${b_i}^2 =
2/a_i + x$. Since $x = 2E/M^{2/3}$ in the hyperbolic case, this result
is equivalent to (\ref{DVsameE.eq}).

The result ${b_i}^2 = 2/a_i - x$ for the elliptic cases, where $x = -
2E/M^{2/3}$, follows in a similar way.

 \section{Calculation of Series Expansions}

   Most of the series expansions given above are
 non-trivial to derive, so a couple of representative calculations are
outlined here.

   In each case, such as equations
 (\ref{SimBangDenHeq})-(\ref{SimBangDenECeq}), it is usually possible to
do a direct series expansion in powers of some suitable small variable,
and then invert it to get a series for $x$.  However, the result is
usually not very tidy, and tends to have sign ambiguities.  Much better
defined and neater results always follow from starting with the parametric
evolution equations,
 (\ref{EllEv})-(\ref{HypEv}).  The calculations were all done to 6th
order, using Maple, but only truncated series are written out here.

   Let us obtain the near-Parabolic series that lies between
(\ref{SimBangDenHeq}) and (\ref{SimBangDenEXeq}).  The parabolic limit
occurs when $x \to 0$, while $R$ and $\tau = t - t_B$ remain finite.  This
requires
 \begin{align}
   \eta \to 0 ~~\mbox{and}~~ \frac{\eta}{\sqrt{x}} \to e
 \end{align}
 so that the new evolution parameter $e$ remains finite for finite $\tau$.
Series expansions of (\ref{HypEv}) then give
 \begin{align}
   \tau & \approx \frac{e^3}{6} + \frac{x e^5}{120} + \frac{x^2 e^7}{5040}
      + \frac{x^3 e^9}{362880} + \cdots ,
      \label{tau-Pseries-e} \\
   a & \approx \frac{e^2}{2} + \frac{x e^4}{24} + \frac{x^2 e^6}{720}
      + \frac{x^3 e^8}{40320} + \cdots .
      \label{a-Pseries-e}
 \end{align}
 Now we invert the above series for $\tau$ by writing
 \begin{align}
   e & \approx e_0 + e_1 x + e_2 x^2 + e_3 x^3 + \cdots ,
 \end{align}
 substituting into (\ref{tau-Pseries-e}), and solving each power of $x$ in
turn for the coefficients $e_i$.  We get
 \begin{align}
   e & \approx (6 \tau)^{1/3} \Bigg( 1
      - \frac{(6 \tau)^{2/3} x}{60} \nn \\
   &~~~~~~ + \frac{(6 \tau)^{4/3} x^2}{1400}
      - \frac{(6 \tau)^2 x^3}{25200} + \cdots \Bigg) ,
      \label{e-Pseries-x}
 \end{align}
 which we substitute into (\ref{a-Pseries-e}),
 \begin{align}
   a & \approx \frac{(6 \tau)^{2/3}}{2} \Bigg( 1
      - \frac{(6 \tau)^{2/3} x}{20} \nn \\
   &~~~~~~ + \frac{3 (6 \tau)^{4/3} x^2}{2800}
      - \frac{23 (6 \tau)^2 x^3}{504000} + \cdots \Bigg) .
      \label{a-Pseries-x}
 \end{align}
 We define $\tau_P$ as the time (since the bang) that it would take an
exactly parabolic model to expand to $a$
 \begin{align}
   a & = \frac{(6 \tau_P)}{2} ,
 \end{align}
 and we define $d\tau_P$ as the difference
 \begin{align}
   d\tau & = \tau - \tau_P .
 \end{align}
 Then we once more invert the series by writing
 \begin{align}
   x & \approx x_1 d\tau + x_2 d\tau^2 + x_3 d\tau^3 + \cdots ,
 \end{align}
 substituting into (\ref{a-Pseries-x}), and solving each power of $d\tau$
in turn for the coefficients $x_i$, thus obtaining
 \begin{align}
   x & \approx \frac{20}{3 a} \left(
      - \frac{d\tau_P}{\tau_P}
      + \frac{25 d\tau_P^2}{14 \tau_P^2}
      - \frac{10000 d\tau_P^3}{3969 \tau_P^3} + \cdots \right) ,
 \end{align}
 which is (\ref{SimBangDenPeq}).

   As our second example we find the near maximum expansion series lying
between (\ref{SimBangVelEXeq}) and (\ref{SimBangVelECeq}).  Maximum
expansion occurs when $\eta \to \pi$, $b \to 0$, $a \to 2/x$, and
$\tau \to \pi/x^{3/2}$.  The mean velocity is
 \begin{align}
   \frac{a_{\text{max}}}{\tau_{\text{max}}}
   = \frac{2}{\pi} \left( \frac{\pi}{\tau_{\text{max}}} \right)^{1/3} .
 \end{align}
 Thus we write
 \begin{align}
   \eta \to \pi + e ,
 \end{align}
 expand in powers of this new $e$,
 \begin{align}
   \tau & \approx \frac{1}{x^{3/2}} \left(
      \pi + 2 e - \frac{e^3}{6} + \cdots \right) ,
      \label{tau-MXseries-e} \\
   b & \approx x \left( \frac{e^2}{4} + \frac{e^4}{24} + \cdots \right) ,
      \label{b-MXseries-e}
 \end{align}
 and invert to get
 \begin{align}
   e & \approx \frac{2}{\sqrt{x}\;} \left( b
      - \frac{b^3}{3 x} + \cdots \right) .
      \label{e-MXseries-b}
 \end{align}
 We then substitute this as well as
 \begin{align}
   x & \approx x_1 b + x_2 b^2 + x_3 b^3 + \cdots ,
 \end{align}
 into (\ref{tau-MXseries-e}), to get the second series inversion:
 \begin{align}
   x & \approx \frac{20}{3 a} \left( 1
      + \frac{16 b}{3 \pi^2 \overline{b}}
      + \frac{64 b^2}{3 \pi^4 \overline{b}^2}
      - \frac{128 (9 \pi^2 - 112) b^3}{81 \pi^6 \overline{b}^3}
      + \cdots \right) ,
 \end{align}
 where
 \begin{align}
   \overline{b} & =
   \frac{2}{\pi} \left( \frac{\pi}{\tau_i} \right)^{1/3}
 \end{align}
 is what the average velocity would be if maximum expansion occurred at
$\tau_i$.  This gives us (\ref{SimBangVelEMeq}).

 \end{document}